\documentclass[12pt,a4paper]{article}
\pdfoutput=1

\usepackage{graphicx,url}
\usepackage{amssymb,amsmath}
\usepackage{epstopdf}
\newtheorem{assumption}{Assumption}

\def\de{{\rm d}}
\def\E{{\rm E}}

\begin{document}

\markboth{De Gregorio and Iacus}
{Clustering Diffusion Processes}


\title{Clustering of discretely observed diffusion processes}

\author{Alessandro De Gregorio\\
{\small Department of Statistics, Probability and Applied Statitics, University of Rome}\\
{\small P.le Aldo Moro 5, 00185 Rome, Italy}\\ 
{\small \url{alessandro.degregorio@uniroma1.it}} \and
Stefano Maria Iacus\\
{\small Department of Economics, Business and Statistics, University of Milan}\\
{\small Via Conservatorio 7, 20124 Milan, Italy}\\
{\small \url{stefano.iacus@unimi.it}}
}
\maketitle


\begin{abstract}
In this paper a new dissimilarity measure to identify groups of assets dynamics is proposed.
The underlying generating process is assumed to be a diffusion process solution of stochastic
differential equations and  observed at discrete time. The mesh of observations is not required to shrink
to zero. As distance between two observed paths, the quadratic
distance of the corresponding estimated Markov operators is considered. 
Analysis of both synthetic data and real financial data
from NYSE/NASDAQ stocks, give evidence that this distance seems capable
to catch differences in both the drift and diffusion coefficients contrary to other commonly
used metrics.
\end{abstract}

{\bf keywords} Clustering of time series; discretely observed diffusion processes; financial assets, markov processes.\\

\section{Introduction}
In recent years, there has been a lot of interest in mining time series data. In particular, financial data are among the most studied data. Although many measures of dissimilarity are available in the literature (see e.g. Liao, 2005, for a review) most of them ignore the underlying structure of the stochastic model which drives the data. Among the few measures which consider the properties of the models we can mention Hirukawa (2006) which considers non-gaussian locally stationary sequences, Corduas and Piccolo (2008) who proposed an autoregressive metric as a distance between ARIMA models, and several information measures  based on the the estimated densities of the processes (see e.g. Kakizawa {\it et. al}, 1998).

Needless to say, starting from the Black and Scholes (1973) and Merton (1973) model, most of models of modern finance rely on continuous time processes. In particular, in most of the cases the dynamic of underlying process used in option pricing is assumed to be a diffusion process solution to some stochastic differential equations. This paper proposes a dissimilarity measure which is particularly taylored to discretely observed diffusion processes. This measure is based on a new application of the results by Hansen {\it et. al} (1998) on identification of diffusion processed observed at discrete time when the time mesh $\Delta$ between observations is not necessarily shrinking to zero. The theory proposed in Hansen {\it et al.} (1998) has been used in Kessler and S\o rensen (1999) and Gobet {\it et al.} (2004) in parametric and non parametric estimation of diffusion processes respectively. The theory is based on the fact that, when the process is not observed at high frequency, i.e. $\Delta\to 0$, the observed data become a true Markov process for which it is possible to identify the Markov operator $P_\Delta$. The continuous time model is instead characterized by the inifinitesimal generator $L_{b,\sigma}$, where $b$ and $\sigma$ are, respectively, the drift and diffusion coefficients of the process. These two operators are equivalent, in the sense of functional analysis, so if one can estimate the  Markov operator from the data it is also possible to identify the process and in particular the couple $(b, \sigma)$. The identification step of this procedure, needs some care (see e.g. Gobet {\it et. al}, 2004). In the present paper, we instead rely on the Markov operator only and use it to build a measure of dissimilarity between two observed processes. Some form of ergodicity or stationarity of the underlying process is usually required although these hypothesis can be relaxed in several directions as, for example, mentioned in Kessler and S\o rensen (1999).

The paper is organized as follows. Section \ref{sec:intro} introduces the model and the assumptions. The Markov operator is presented in Section \ref{sec:markov}. Section \ref{sec:analysis} studies the performance of the method. First, the behaviour of the operator is analyzed  on simulated paths when data belong to the same hypothetical groups. Finally, real data from the NYSE/NASDAQ are analyzed. All the results include a comparison with other three dissimilarity measures, namely, the Euclidean distance, the  short-time series distance and the dynamic time warping distance. All plots and figures are contained after the references in Section \ref{sec:fig}.

\section{Model and assumptions}\label{sec:intro}
Let $I=(l,r),-\infty \leqslant l < r \leqslant +\infty$ be the state space of a time-homogeneous diffusion process $\{X_t, t\geqslant 0\}$  solution of a stochastic differential of the form
\begin{equation}
\de X_t = b(X_t) \de t + \sigma(X_t) \de W_t
\label{eq:sde1}
\end{equation}
In the expression \eqref{eq:sde1}, $b:I\to \mathbb{R}$ and $\sigma:I\to(0,\infty)$ represent drift and diffusion coefficient, while $W_t$ is a standard brownian motion.
\begin{assumption}\label{ass:exun}
The drift and diffusion coefficient are such that the stochastic differential equation \eqref{eq:sde1} admits a unique weak solution $X_t$.
\end{assumption}
Let us introduce  the scale function and speed measure, defined respectively as
\begin{equation}
s(x) = \exp\left\{-2\int_{\tilde x}^x\frac{b(y)}{\sigma^2(y)}\de y\right\}\,,
\label{eq:scalem}
\end{equation}
with $\tilde x$ any value in the state space $(l,r)$,
and
\begin{equation}
m(x) = \frac{1}{\sigma^2(x)s(x)}\,.
\label{eq:speedm}
\end{equation}

\begin{assumption}\label{ass:ergodic}
We assume that
$$C_0 = \int_{l}^{r} m(x) \de x < \infty\,.
$$
Let, $x^*$ be  an arbitrary point  in the state space of $X_t$ such that 
$$
\int_{x^*}^{r} s(x) \de x =+\infty,\, \int_{l}^{x^*} s(x) \de x = -\infty\,.
$$
If one or both of the above integrals are finite, the corresponding boundary is assumed to be instantaneously  reflecting. 
\end{assumption}

If the Assumption \ref{ass:exun}-\ref{ass:ergodic} are satisfied, then exists a unique ergodic  process $X_t$ solution for the stochastic differential equation \eqref{eq:sde1}, with invariant law 
\begin{equation}
\mu_{b,\sigma}(x) = \frac{m(x)}{C_0} =  \frac{ \exp\left\{2\int_{\tilde x}^{x} \frac{ b(y)}{\sigma^2(y)} \de y\right\}}{C_0\sigma^2(x)}
\label{eq:invdens}
\end{equation}
\section{The Markov operator}\label{sec:markov}
Consider now the regularly sampled data $X_i = X(i \Delta)$, $i=0, \ldots, N$, from the sample path of $X_t$, where $\Delta>0$ and not shrinking to 0 and such that $T=N\Delta$. The process $\{X_i\}_{i=0, \ldots, N}$ is a Markov process and under mild regularity conditions, all the mathematical properties are embodied in the transition operator
$$P_\Delta f(x) = \E\{ f(X_i) | X_{i-1} = x\}\,.$$
Notice that $P_\Delta$ depends on the transition density between $X_i $ and $X_{i-1}$, so we put explicitly the dependence on $\Delta$ in the notation.
This operator is associated with the infinitesimal generator of the diffusion $L_{b,\sigma}$ which is the following operator on the space of continuous and twice differentiable functions
$$
L_{b,\sigma} f(x) = \frac{\sigma^2(x)}{2} f''(x) + b(x)f'(x)\,.
$$
When the invariant density $\mu=\mu_{b,\sigma}(\cdot)$ of the process $X_t$ exists, the operator is unbounded but self-adjoint negative on $L^2(\mu) = \{ f : \int |f|^2\de\mu<\infty\}$ and the functional calculus gives the correspondence (in terms of operator notation)
\begin{equation}
P_\Delta = \exp\{\Delta L_{\mu}\}
\label{eq:markov}
\end{equation}
This relation has been first noticed by Hansen {\it et al.} (1998) and Chen {\it et al.} (1997). It was then used in statistics to derive estimating functions based on the eigenvalues of the above problem by Kessler and S\o rensen (1999). Indeed, to estimate parametrically the coefficients $\sigma(x)=\sigma_\theta(x)$ and $b=b_\theta(x)$ of \eqref{eq:sde1} it suffices to notice  that
$$L_\theta f(x) = \frac{\sigma^2_\theta(x)}{2} f''(x) + b_\theta(x) f'(x)
$$ can be seen as an eigenvalue problem
$L_\theta \psi_\theta(x) = \kappa_\theta\psi_\theta(x)$ and the pair $(\kappa_\theta, \psi_\theta)$ satisfies
$$
P_\Delta \psi_\theta(X_i) = \E\{ \psi_\theta(X_{i+1}) | X_i\} = \exp(\kappa_\theta \Delta)\psi_\theta(X_i)\,.
$$
When the solution is available it is then possible to impose a set of moment condition from which estimating functions are obtained. More recently, under low sampling rate, the result \eqref{eq:markov} was used to estimate non parametrically the drift and diffusion coefficient  by Gobet {\it et al.} (2004). Indeed, consider the explicit form of the invariant law $\mu_{b,\sigma}$
in \eqref{eq:invdens} and define
 $S(x) = 1/s'(x) = \frac12\sigma^2(x)/\mu_{b,\sigma}(x)$ (see also A\"it-Sahalia, 1996). Being $\nu_1$ the largest negative eigenvalue of $L_{\mu,\sigma}$, the following eigenvalue problem can be written
$$L_{b,\sigma} u_1(x) = \frac{1}{\mu_{b,\sigma}(x)}\left(S(x) u_1'(x) \right)'= \nu_1 u_1(x)$$
from which $S(x)u_1'(x) = \nu_1\int_{l}^x u_1(y)\mu_{b,\sigma}(y) \de y$. Finally,
\begin{equation}
\sigma^2(x) = \frac{2\nu_1\int_{l}^x u_1(y)\mu_{b,\sigma}(y)\de y}{u_1'(x)\mu_{b,\sigma}(x)}
\label{eq:diff}
\end{equation}
and
\begin{equation}
b(x) = \nu_1 \frac{u_1(x)u_1'(x)\mu_{b,\sigma}(x)-u_1''(x) \int_{l}^x u_1(y)\mu_{b,\sigma}(y)\de y}{u_1'(x)^2\mu_{b,\sigma}(x)}
\label{eq:drift}
\end{equation}
When $P_\Delta$ can be estimated properly from the data, the pair $(u_1, \nu_1)$ can be obtained as well and then plugging these values into the above expressions \eqref{eq:drift} and \eqref{eq:diff} estimators of $b(\cdot)$ and $\sigma(\cdot)$  are obtained.

In this paper, we propose to use an the estimator of $P_\Delta$ and from this build a distance between discretely observed diffusion processes.

For a given $L^2$-orthonormal basis $\{\phi_j, j\in J\}$ of $L^2([l, r])$, where $J$ is an index set, following Gobet {\it et. al} (2004) it is possible to obtain an estimator $\hat {\bf P}_\Delta$ of $<P_\Delta \phi_j,\phi_k>_{\mu_{b,\sigma}}$ with entries
\begin{equation}
(\hat P_\Delta)_{j, k}(X)  = \frac{1}{2N} \sum_{i=1}^N \left\{\phi_j(X_{i-1})\phi_k(X_i) + \phi_k(X_{i-1})\phi_j(X_i)\right\}, \quad j,k\in J
\label{eq:MO}
\end{equation}
The terms $(\hat P_\Delta)_{j, k}$ are approximations of 
$<P_\Delta \phi_j, \phi_k>_{\mu_{b,\sigma}}$, that is, the action of the transition operator on the state space with respect of the unknown scalar product $<\cdot,\cdot>_{\mu_{b,\sigma}}$ and hence can be used as ``proxy''  of the probability structure of the model.

We remark that, like the invariant density $\mu_{b,\sigma}$, the Markov operator itself cannot perfectly identify the underlying process, in the sense that, for some $(b_1, \sigma_1)$ there might exist another couple $(b_2, \sigma_2)$ such that $\mu_{b_1, \sigma_1}(x) = \mu_{b_2, \sigma_2}(x)$ and the same applies to the infinitesimal generator and hence to the Markov operator. So in this sense, the identification cannot be precise: unicity is not guaranteed. Nevertheless, the measure proposed in the next section, can only help in finding similarities of two (or more) processes in terms of the action of the Markov operator on the approximating space generated by the basis of $L^2$ above. Indeed, the Markov operator also takes into account the transition properties of the observed sequence which is the natural way to make inference from discretely observed diffusion processes.

\section{Analysis of  performance of the proposed method}\label{sec:analysis}
In this section we consider four different distances to be used in both the analysis of synthetic data  and on real financial time series.  For an updated review on time series dissimilarity measures see Liao (2005). In the following, we will denote by $X = \{X_i, i=1, \ldots, N\}$ and $Y = \{Y_i, i=1, \ldots, N\}$ two generic paths. 
We will consider the following measures

\paragraph{The Markov-Operator distance}
Following the suggestion in Rei\ss (2003) we use a basis of 20 orthonormal B-splines on a compact support (see Ramsay and Silverman, 2005) of degree 10. As compact support we consider the observed support of all diffusion paths enlarged by 10\%.
In the analysis of synthetic data, the support is just the interval [0,1].
Then we define the Markov Operator distance as follows
$$
d_{MO}(X,Y) = \sum_{j,k\in J} [(\hat P_\Delta)_{j, k}(X) - (\hat P_\Delta)_{j, k}(Y)]^2 
$$
where $(\hat P_\Delta)_{j, k}(X)$ is calculated as in \eqref{eq:MO}. 

\paragraph{Short-Time-Series distance}
Proposed by M\"oller-Levet {\it et al.} (2001) is based on the idea to consider each time series as a piecewise linear function and compare the slopes between all the interpolants. It reads as
$$
d_{STS}(X, Y) = \sqrt{\sum_{i=1}^N \left(
\frac{X_{i} - X_{i-1}}{\Delta}-
\frac{Y_{i} - Y_{i-1}}{\Delta}
\right)^2}
$$
This measure is essentially design to discover similarities in the volatility between two time series regarding of the average level of the process, i.e. one process and a shifted version of it will have zero distance.

\paragraph{The Euclidean distance}
The usual Euclidean distance is one of the most used in the applied literature, in particular in one step ahead prediction. We will calculate it as follows
$$
d_{EUC}(X, Y) = \sqrt{\sum_{i=1}^N \left(
X_{i} - Y_{i}
\right)^2}
$$
and use only for comparison purposes.

\paragraph{Dynamic Time Warping distance}
The Euclidean distance is very sensitive to distortion in time axis and may lead to poor results for sequences which are similar, but locally out of phase (Corduas, 2007).The Dynamic Time Warping (DTW) was introduced originally in speech recognition analysis (Sakoe and Ciba, 1978; Wang and Gasser, 1997). DTW allows for non-linear alignments between time series not necessarily of the same length. Essentially,  all shiftings between two time series are attempted and each time a cost function is applied (e.g. a weighted Euclidean distance between the shifted series). The minimum of the cost function over all possible shifting is the dynamic warping distance $d_{DTW}$. In our applications we use the Euclidean distance in the cost function and the algorithm as implemented in the R package $\tt dtw$ (Giorgino, 2007).

\subsection{Analysis of synthetic data}\label{sec:synth}
We simulate 10 paths $X_i$, $i=1, \ldots, 10$, according to the combinations of drift $b_i$ and diffusion coefficients  $\sigma_i$, $i=1, \ldots, 4$ presented in the following table
\begin{center}
\begin{tabular}{c|cccc}
& $\sigma_1(x)$ & $\sigma_2(x)$ & $\sigma_3(x) $ & $\sigma_4(x) $\\ 
\hline
$b_1(x)$ & X10, X1 & &X5&\\
$b_2(x)$& & X2,X3 & X4\\ 
$b_3(x)$ &&X6, X7\\
$b_4(x)$ & & & &X8\\
\end{tabular}
\end{center}
where
$$
b_1(x) = 1-2x, \quad
b_2(x) = 1.5(0.9-x),\quad 
b_3(x) = 1.5(0.5-x),\quad
b_4(x) = 5(0.05-x)\quad
$$
and
$$\sigma_1(x) = 0.5+2 x (1-x),\quad \sigma_2(x) = \sqrt{0.55 x (1-x)}$$
$$ \sigma_3(x) = \sqrt{0.1 x (1-x)},\quad \sigma_4(x) = \sqrt{0.8 x (1-x)}$$
The process $X9=1-X1$, hence it has drift $-b_1(x)$ and the same quadratic variation of $X1$ and $X10$.

\begin{figure}[h]
\includegraphics[width=\textwidth]{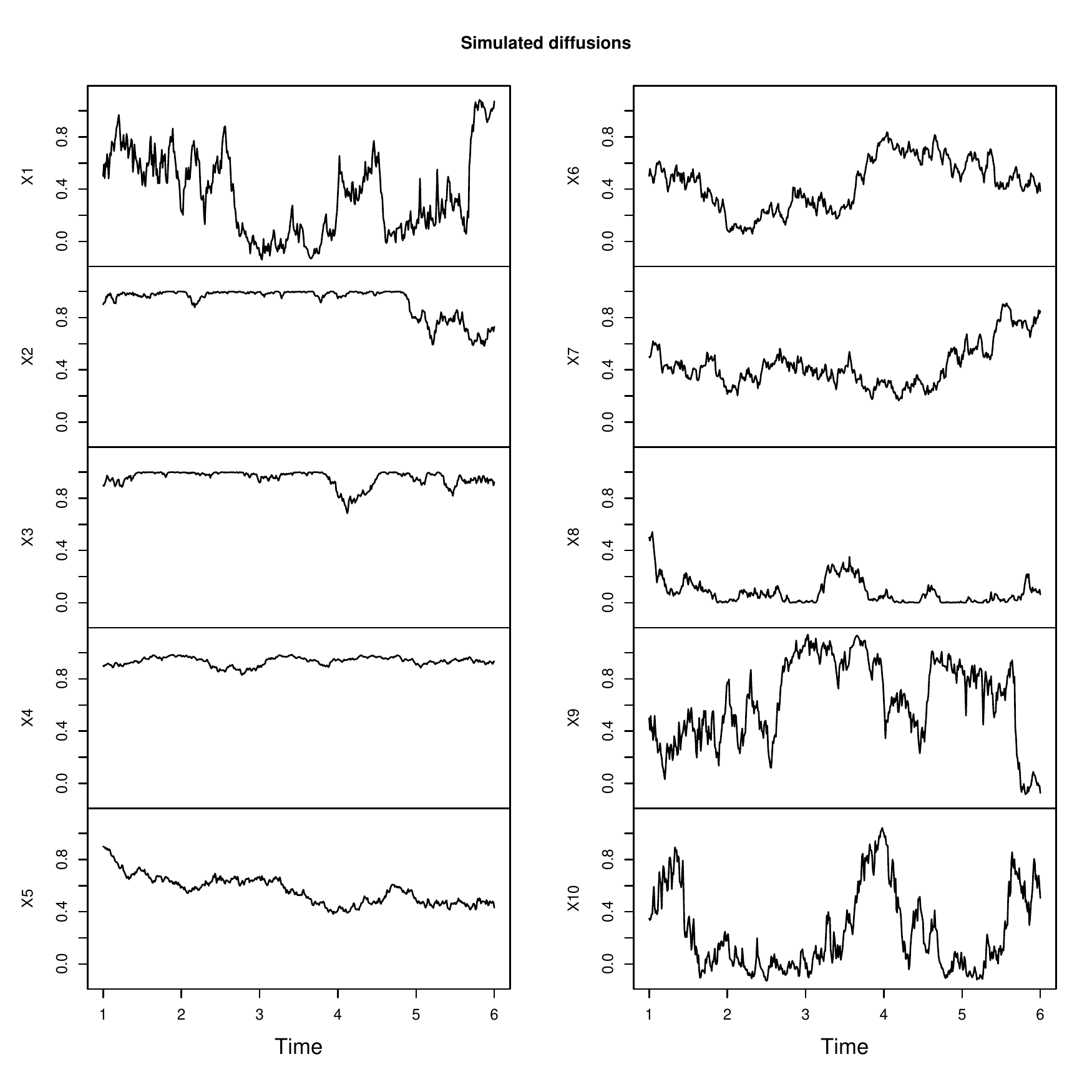}
\caption{Paths of different processes simulated according to Section \ref{sec:synth}.}
\label{fig:series}
\end{figure}

We simulate each path using (second) Milstein scheme (see e.g. Kloden {\it et al.} 1999 or Iacus, 2008) with time lag $\delta=1e-4$ each path of length 50,000. Observations have been then  resampled at rate $\Delta=0.1$, so the observed path used in the estimation have length $N=500$. The sample path of process X9 is a reflection of the sample path of X1 around 1, i.e. X9 = 1 - X1. The final paths are reported in Figure \ref{fig:series}

\begin{figure}[h]
\includegraphics[width=\textwidth]{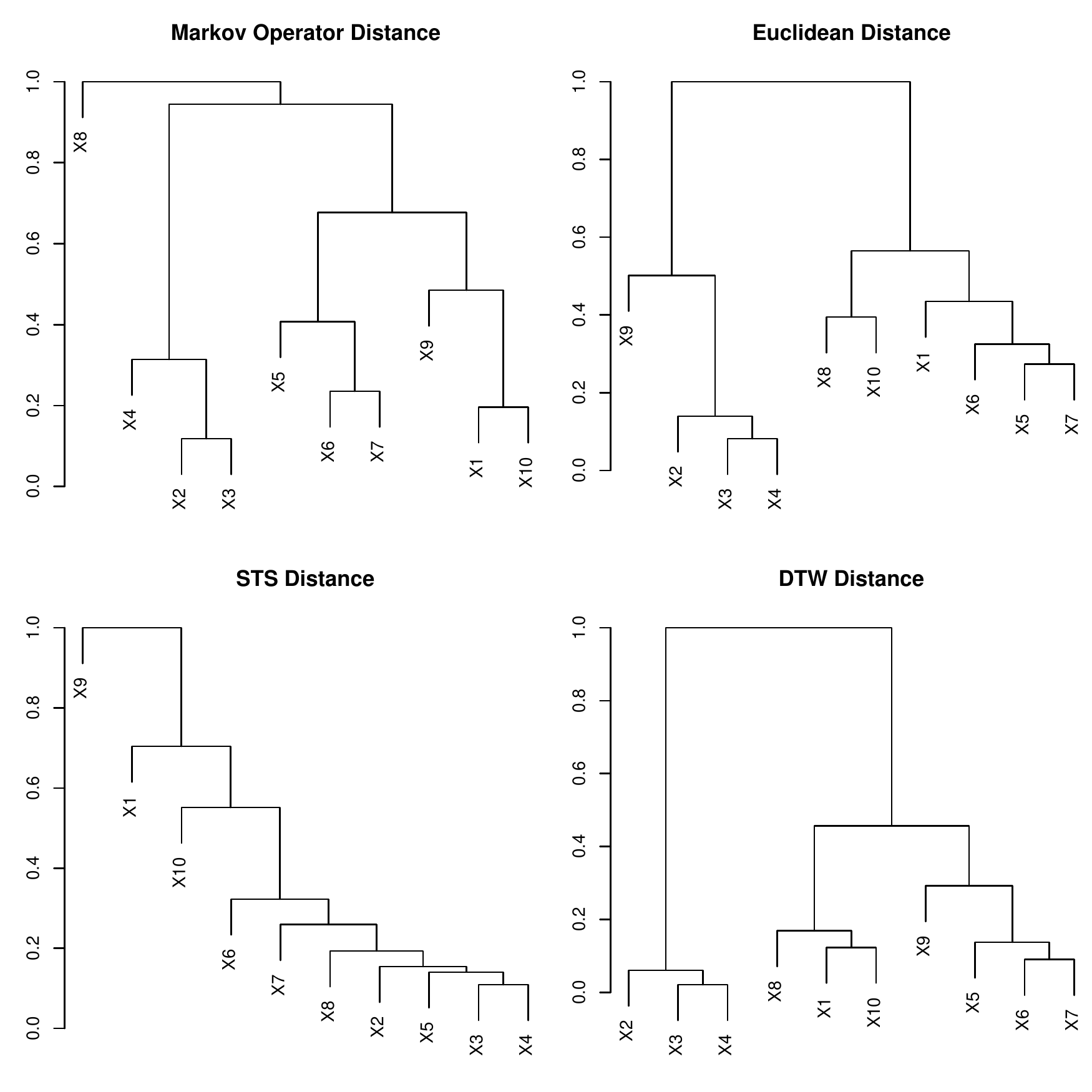}
\caption{Clustering according to different distances.}
\label{fig:dendro}
\end{figure}

After applying the distance $d_{MO}$, $d_{STS}$, $d_{EUC}$ and $d_{DTW}$ we run hierarchical clustering with complete linkage method. To make the output graphically comparable we rescale all distance matrixes to (0,1). This rescaling only gives the feeling of relative distance of observations, so numerical values are not really comparable from one distance to another.
Figure \ref{fig:dendro} shows the final classification. From the plots it appears quite evident, that apart from scaling, that $d_{STS}$ and $d_{EUC}$ agrees but unfortunately in this example, do not correctly classify the paths. 
The processes X1 and X10 are driven by the same stochastic differential equation although their initial values are different, while this difference increases the distances $d_{STS}$ and $d_{EUC}$ the Markov operator distance 
$d_{MO}$ seems to correctly catch the similar effect of drift and diffusion coefficient on the path. Similarly for X9 which is just a reflection of X1 around 1. In Figure \ref{fig:dendro} it easy to see that while $d_{MO}$ puts X1 and X10 together and this cluster together with X9, the other two distance put X9 in a separate cluster which is then aggregated with  trajectories in different clusters not really related to X9 in terms of drift and diffusion coefficients.
Processes X2 and X3 are also driven by the same stochastic differential equation and this is captured also by the other two distances although $d_{MO}$ put X2 and X3 in the smaller cluster and then aggregate with X4 which has the same drift as X2 and X3 but a different diffusion coefficient. The other two methods put X3 and X4 together and then aggregate X2.
A similar situation occurs for X6 and X7 which have the same stochastic differential equations which clearly separated by $d_{MO}$ and not for the other two distances. Finally, $d_{MO}$ clearly separates X8 which is the real outlier in terms of drift and diffusion coefficient.
 This fact is not captured by the other measures.

\subsection{Analysis of real financial data}
\begin{figure}[h]
\includegraphics[width=\textwidth]{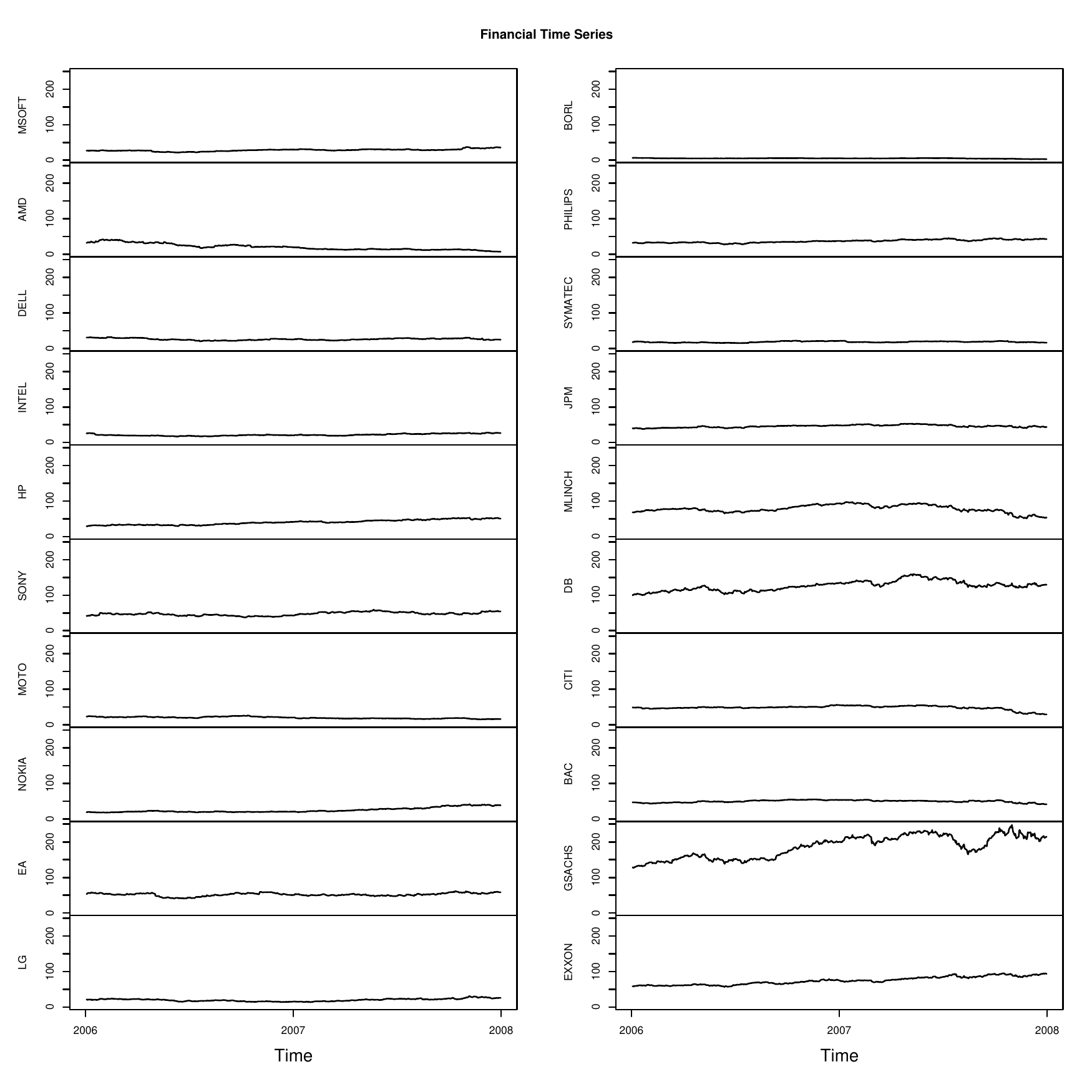}
\caption{Paths of the 20 assets considered: from 2006-01-03 to 2007-12-31.}
\label{fig:series2}
\end{figure}

\begin{figure}[h]
\includegraphics[width=\textwidth]{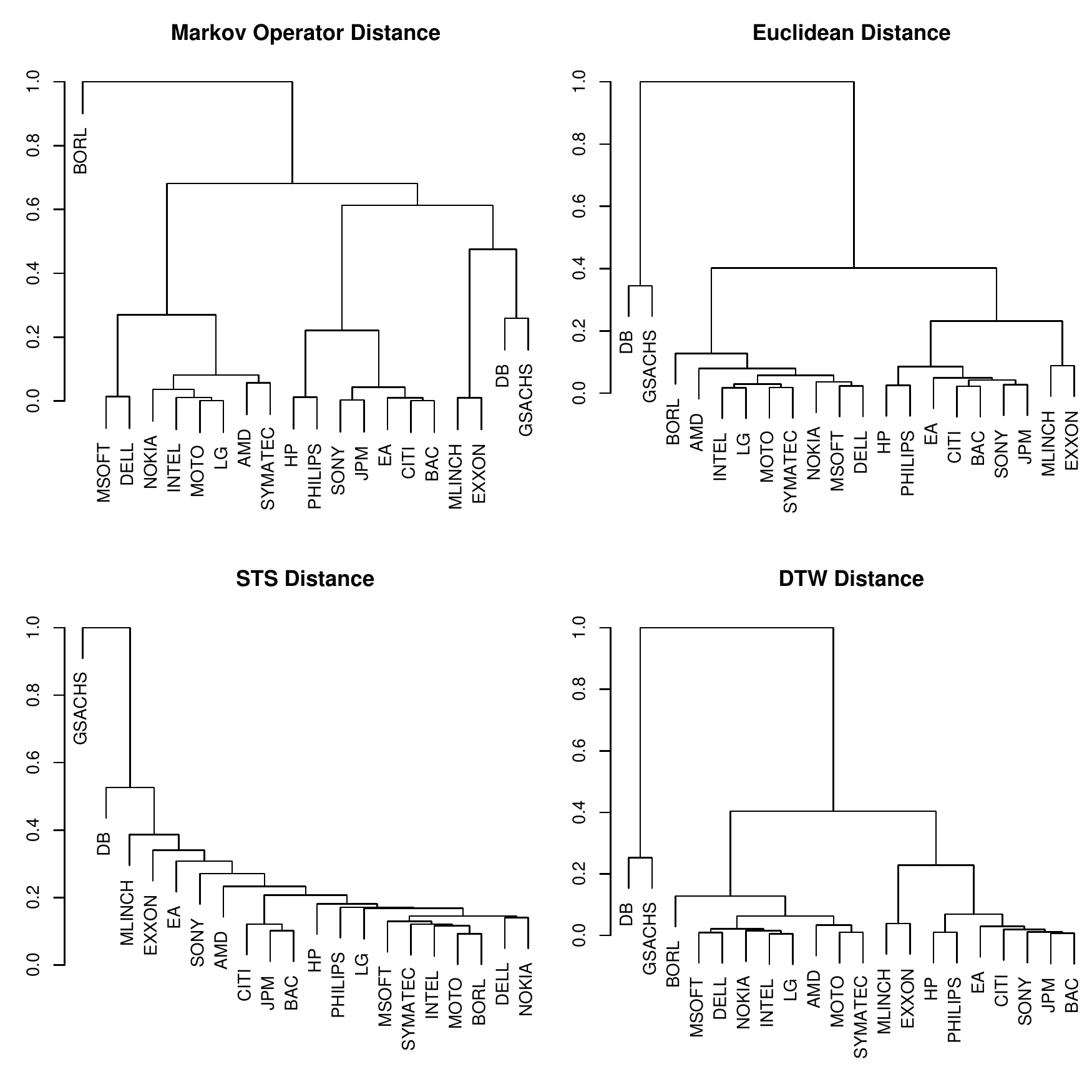}
\caption{Clustering according to different distances.}
\label{fig:dendro2}
\end{figure}

\begin{figure}[h]
\includegraphics[width=\textwidth]{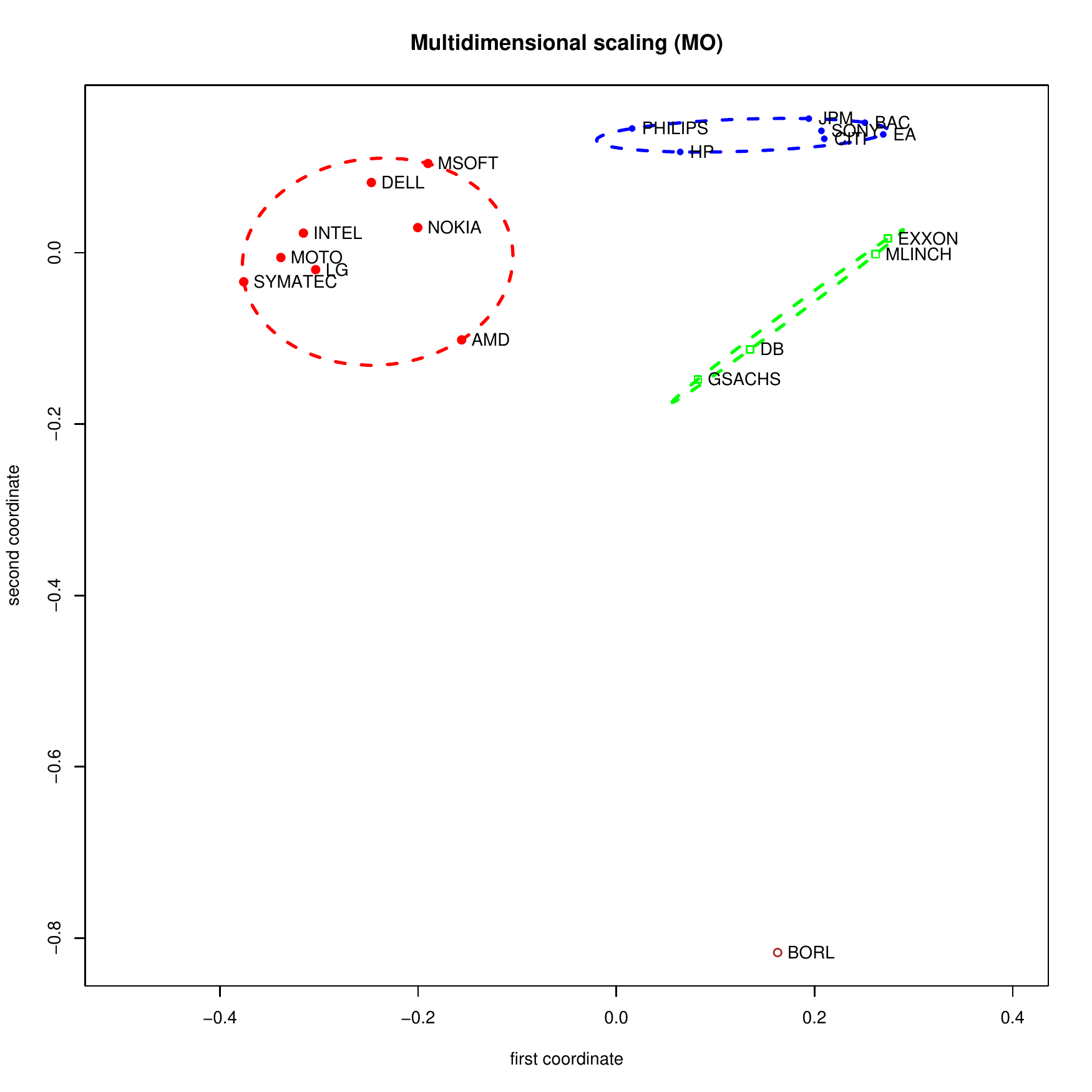}
\caption{Multidimensional scaling on the distance matrix $d_{MO}$. Observations in the same cluster are plotted with the same symbol. If a cluster contains more than two observation, the ellipsoid hull is also represented.}
\label{fig:mds2}
\end{figure}

We consider the time series of daily closing quotes, from 2006-01-03 to 2007-12-31, for the following 20 financial assets:
Microsoft Corporation (MSOFT in the plots), Advanced Micro Devices Inc. (AMD), Dell Inc. (DELL),
Intel Corporation (INTEL), Hewlett-Packard Co. (HP), Sony Corp. (SONY), 
Motorola Inc. (MOTO), Nokia Corp. (NOKIA), Electronic Arts Inc. (EA), 
LG Display Co., Ltd. (LG), Borland Software Corp. (BORL), 
Koninklijke Philips Electronics NV (PHILIPS), 
Symantec Corporation (SYMATEC),
JPMorgan Chase \& Co (JMP),
Merrill Lynch \& Co., Inc. (MLINCH),
Deutsche Bank AG (DB), Citigroup Inc. (CITI),
Bank of America Corporation (BAC),
Goldman Sachs Group Inc. (GSACHS) and
Exxon Mobil Corp. (EXXON). Quotes come from NYSE/NASDAQ. Source Yahoo.com. Missing values have been linearly interpolated.
These assets come from both electronic hardware, appliance and software vendors or producers, financial institutions of different type and a petrol company. 
Figure \ref{fig:series2} represents the 20 paths of the assets all on the same scale in order to make them comparable in a visual inspection. It is clear that some titles have larger volatility than others and possibly there some outlier in terms of both trend and volatility. For example, looking at financial companies, one can notice that MLINCH, DB amd GSACHS, although at different volatility levels, all present the same (cyclic) drift behaviour over time. Further, CIT and BAC seems quite close in terms of volatility and drift. But visual inspection alone is not sufficient so try to discover clusters using the for distances introduced before.
Figure \ref{fig:dendro2} reports the four different dendrograms for the four metrics.
While all methods seems to separate DB and GSACHS, only $d_{MO}$ seems to collect most of financial companies in the same parent cluster. Our metrics clearly separates BORL (as an outlier or singleton) in a cluster very far form the other observed paths. Also $d_{TW}$ and $d_{EUC}$ tend to separate BORL as well, but the identified cluster is closer to other observations than other clusters. The metric $d_{STS}$ does not appear to give sharp indication on how to separate clusters.
Indeed, if we decide to split the dendrogram into  four clusters, $d_{MO}$ separates BORL in one cluster, DB, GSACHS, MLINCH and EXXON in the secod cluster, a group of hardware producer (mostly) and a final group of less active financial assets (CITI, JPM) and appliance producers  or hardware assemblers (SONY, PHILIPS, HP). EA goes together with SONY in all dendrograms, which is not an unrealistic evidence in that the company essentially produces software for game consoles.
Figure \ref{fig:mds2} present the multidimensional scaling of the distance matrix in which groups identified by the cluster are plotted with the same symbol and, when cluster contain more than two elements, the ellipsoid hull is also drawn.

The present very superficial analysis of the clustering should not go in depth with financial implications of the results. Nevertheless, the conclusion of the analysis is that, although all metrics have pros and cons because they look at single different aspects, the Markov operator distance seems able to discriminate discrepancies in both volatility and drift of the observed processes. It also give a sharp indication on where to cut the dendrogram to obtain comparable groups.


\newpage

\section{Figures}\label{sec:fig}

\begin{figure}[h]
\includegraphics[width=\textwidth]{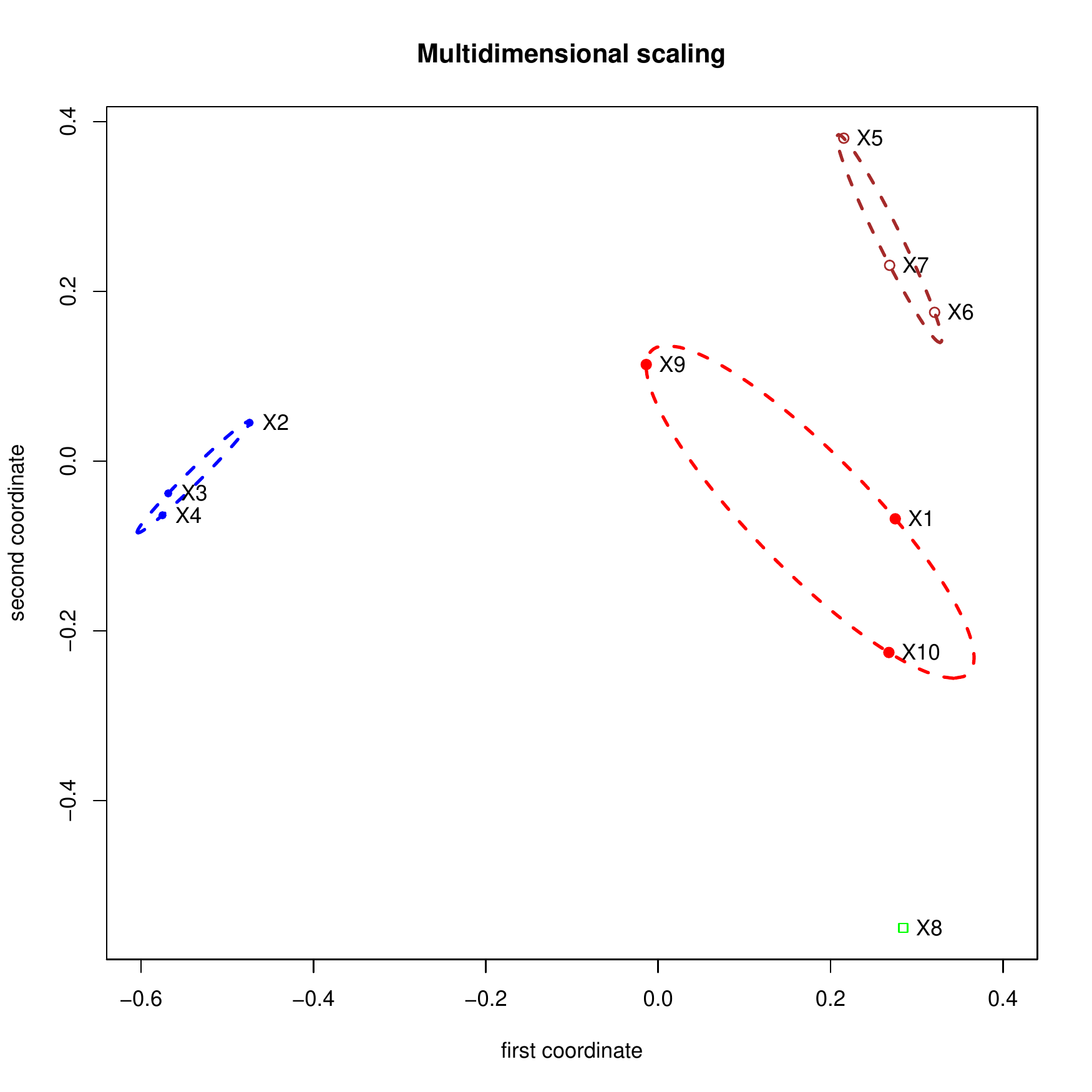}
\caption{Multidimensional scaling representation of distance $d_{MO}$ with points identified after cutting dendrogram 1 in Figure \ref{fig:dendro} into  4 groups.}
\label{fig:mds}
\end{figure}

\end{document}